\documentclass[journal]{IEEEtran}
\pdfoutput=1
\usepackage{cite}
\usepackage{moreverb}
\usepackage{tabularx}
\usepackage{pifont}
\usepackage{multirow}
\usepackage[center]{caption}

\usepackage{graphicx}
\usepackage{amsmath}

\usepackage{algorithmic}

\usepackage{array}

\newcommand{\etal}{\textit{et al.}}

\begin{document}
\title{Federated learning and next generation wireless communications: A survey on bidirectional relationship}

\author{Debaditya Shome,
        Omer Waqar and
        Wali Ullah Khan
\thanks{Debaditya Shome was with the Department of Engineering, Thompson Rivers University, BC, Canada.}
\thanks{Omer Waqar is with the Department of Engineering, Thompson Rivers University (TRU), BC, Canada.}
\thanks{Wali Ullah Khan is with the Interdisciplinary Center for Security, Reliability and Trust (SnT), University of Luxembourg, 1855 Luxembourg City, Luxembourg.

This work is supported through funding from the Mitacs Canada under the Mitacs Globalink research internship program.}

}

%
%

\markboth{Accepted in Transactions on Emerging Telecommunications Technologies (Wiley)}%
{Shome \MakeLowercase{\textit{et al.}}}

\maketitle

\begin{abstract}
In order to meet the extremely heterogeneous requirements of the next generation wireless communication networks, research community is increasingly dependent on using machine learning solutions for real-time decision-making and radio resource management. Traditional machine learning employs fully centralized architecture in which the entire training data is collected at one node e.g., cloud server, that significantly increases the communication overheads and also raises severe privacy concerns. Towards this end, a distributed machine learning paradigm termed as Federated learning (FL) has been proposed recently. In FL, each participating edge device trains its local model by using its own training data. Then, via the wireless channels the weights or parameters of the locally trained models are sent to the central parameter server (PS), that aggregates them and updates the global model. On one hand, FL plays an important role for optimizing the resources of wireless communication networks, on the other hand, wireless communications is crucial for FL. Thus, a `bidirectional' relationship exists between FL and wireless communications. Although FL is an emerging concept, many publications have already been published in the domain of FL and its applications for next generation wireless networks. Nevertheless, we noticed that none of the works have highlighted the bidirectional relationship between FL and wireless communications. Therefore, the purpose of this survey paper is to bridge this gap in literature by providing a timely and comprehensive discussion on the interdependency between FL and wireless communications.
\end{abstract}


%
\IEEEpeerreviewmaketitle

\section{Introduction}\label{sec1}
The wireless communication networks are experiencing a paradigm shift from being smartphone-centric to become more of an Internet-of-Things (IoT) oriented system that connects billions of miniature and power-limited devices \cite{8755300}. In order to support hyper-connectivity of these massive IoT networks, the next generation wireless communication systems such as sixth generation (6G) networks are envisioned to be characterized by an unprecedented flexible, adaptive and reconfigurable architectures. Hence, the next generation communication systems will entail optimization of the sheer number of network parameters. Keeping this fact in mind, it is clear that existing optimization approaches that are based on conventional mathematical models will either become completely obsolete or highly inadequate for the next generation wireless communication systems. In other words, it will be the first time since the inception of wireless networks that a \textit{complexity crunch} will exist for
optimizing the resource allocations of the next generation wireless networks. Moreover, the current non-data driven techniques will be unable to achieve autonomy in the next generation of wireless networks. This necessitates to develop new data-driven techniques which either complement or completely replace the traditional non data-driven optimization approaches and should be able to orchestrate the network resources in an intelligent manner.
\begin{table*}[t]
    \caption{Survey articles on FL in the domain of wireless communications.} \label{tab:1}
        \centering
        \begin{tabular}{|c|c|c|c|}
            \hline
            \multirow{1}{*}{\textbf{Title}} &
            \multirow{1}{*}{\textbf{FL for wireless}} &
            \multirow{1}{*}{\textbf{Wireless for FL}} &
            \multirow{1}{*}{\textbf{Communication efficient FL}} \\ 
            \hline
            Niknam \etal\cite{niknam2020federated}   & \ding{52} &  & \\ 
            Lim \etal \cite{lim2020federated} & \ding{52} &   & \ding{52} \\ 
            Liu \etal \cite{9205981}    & \ding{52} &     & \ding{52} \\
            Khan \etal \cite{khan2020federated}    &\ding{52}  &     & \ding{52} \\ 
            Wahab \etal \cite{wahab2021federated} & \ding{52} &  &  \ding{52} \\
            Xia \etal \cite{xia2021survey} & \ding{52} & & \ding{52} \\
            Du \etal \cite{du2020federated} & \ding{52} & & \\
            Chen \etal\cite{chen2021distributed} & & \ding{52} & \ding{52} \\
            Gafni \etal\cite{gafni2021federated} & & \ding{52} & \\ 
            Hu \etal\cite{hu2021distributed}  &\ding{52}  &     & \ding{52} \\
            Hellstrom \etal\cite{hellstrom2020wireless} & & \ding{52} & \\
            \textbf{Our paper}    & \ding{52} & \ding{52}    & \ding{52}\\
           \hline
        \end{tabular}
\end{table*}
The performance of deep-learning has been phenomenal in areas such as computer vision, natural language processing (NLP) and speech recognition etc\cite{dong2021survey, dargan2020survey}. Moreover, the performance of deep learning is surpassing the performance of ordinary machine learning algorithms with the ever-increasing amount of trainable data and the recent advancements in computing devices e.g., Graphical Processing Units (GPU)\cite{chen2019deep}. Motivated by this fact and  to address the aforementioned complexity crunch challenge, researchers have shown interest to develop new deep learning based solutions specifically tailored for next generation wireless communication systems.
The job of a typical resource management unit in any wireless system is to allocate network resources (e.g., transmission power, bandwidth, beamforming vector, time slots, antennas, base-stations etc.) with an objective to optimize one or more performance metrics (e.g., throughput). In case, an accurate mathematical model exists that characterizes the performance metric, then any resource allocation decision can be cast into an optimization problem. Apparently, solving such an optimization problem seems to be a trivial task. However, considering the fact that the next  generation wireless communication systems will be highly complex in nature owing to integration of advanced technologies such as highly dense small cell networks, Massive multiple-input-multiple-output (MIMO), distributed MIMO and Intelligent reflecting surfaces (IRS) etc., the underlying optimization problems are generally non-deterministic polynomial time (NP)-hard. Therefore, the complexity for solving these problems increases exponentially with the number of radio parameters which are far more in number for future networks as compared to the current wireless networks. Furthermore, the resource allocation decision has to be repeated every time when any of the system parameters (e.g., users’ position, channel realizations etc.) change its value and this change occurs quite frequently particularly in high mobility environments due to shorter channel coherence time. Therefore, the real-time implementation of existing resource allocation (or optimization) algorithms is a huge challenge for highly complex next generation wireless communication systems. However, the use of deep learning techniques with a trained neural network  will enable true online (real-time) resource allocation and thus solve the NP-hard problems efficiently. Moreover, it
is well-known that neural networks are universal function approximators\cite{hornik1989multilayer}, thus it is quite possible to either completely eliminate or reduce the dependence of human knowledge and to achieve complete autonomy for the wireless networks.

Majority of the deep learning related works in many fields including computer vision, NLP and wireless communications \cite{voulodimos2018deep, otter2020survey, zhang2019deep} considered a central entity that accommodates an entire raw data for training the neural network. As discussed above, these deep learning based approaches have shown remarkable performance for solving challenging non-convex optimization problems, however some critical issues have been observed in the centralized architecture of these deep-learning solutions. Firstly, a massive amount of data needs to be stored, processed and trained at the central entity, which seems to be impractical as the training process on a single compute instance is very time-consuming \cite{9120226}. Secondly, as the raw data has to be uploaded to the central entity, this uplink data offloading can be extremely expensive for devices with short battery life, such as smartphones and IoT nodes. Thirdly, uploading such large volume of raw data exacerbates the stress on wireless networks and can cause network congestion. Fourthly, devices that are used for privacy-sensitive applications, owing to the potential of theft of their valuable data,  may be unwilling to upload their raw data\cite{hosseinalipour2020federated}.

Federated learning (FL) has recently emerged as a disruptive distributed learning paradigm that has an ability to overcome all the aforementioned issues of the centralized learning approaches. Moreover, FL facilitates Artificial Intelligence (AI) democratization in which besides the large enterprises, small organisations can also play role in data storage, processing, inference and decision-making. It is envisaged that 6G communication systems will be based on the concept of self-sustaining networks, which refer to the networks that can optimize themselves automatically and handle all dynamic interactions with the  real-world environment without human involvement. Traditional centralized deep learning algorithms will not be able to achieve the desired outcomes due to tremendous computational overload in these dynamic network settings, which create massive volumes of data in a short amount of time. Therefore,  we believe that FL will be a critical component in reaching the objective of self-sustaining networks, as managing massive dynamic data will not be a problem when there is a large number of devices available to store, process, compute such huge data and to train a single global model collaboratively.\\

\section{Related surveys on FL and Our Contributions}

\begin{figure*}[t!]
\centering
\includegraphics[width=0.75\textwidth, height=5in]{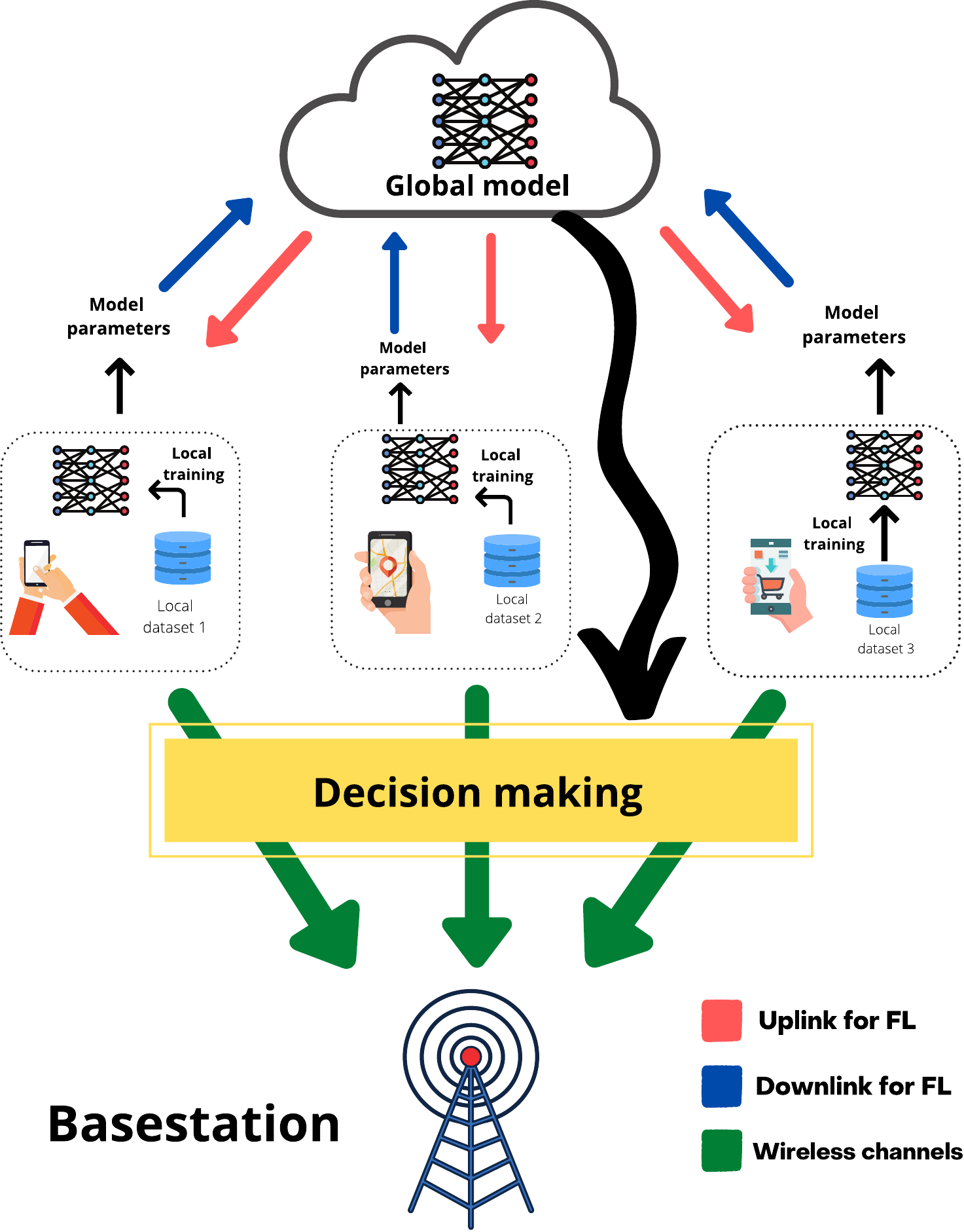}
\caption{Illustration of the Bi-directional relationship}
\label{fig:1}
\end{figure*}

Motivated by the huge advantages of the FL when compared to the centralized learning approaches, recently few survey articles have appeared that discuss FL in the domain of wireless communications. These survey papers are listed in Table 1.  In \cite{niknam2020federated}, the authors survey FL and its applications towards 5G networks along with the future directions. Lim \etal \cite{lim2020federated} present a detailed tutorial on FL and a comprehensive survey on challenges and applications of FL in mobile edge networks (MEC). It is worth mentioning here that this paper put more focus on the statistical challenges of FL which have been less discussed in prior literature. Liu \etal \cite{9205981}, provide a detailed survey on the impact of FL for achieving the goals of 6G communication systems. This paper discusses the core challenges in the deployment of FL in wireless networks and the authors propose some key ideas to solve these challenges.  
In context of the IoT networks, Khan \etal \cite{khan2020federated} provide a survey that discusses the advances, evaluation metrics, challenges, taxonomy of literature and future directions of FL . Furthermore, the authors in \cite{wahab2021federated} present a detailed survey on FL for the advancements of wireless communications with focus on grouping the existing literature using a multi-level classification scheme. Furthermore, Xia \etal\cite{xia2021survey} provide a detailed analysis on the intersection of FL and edge computing with a focus on the challenges, applications, security, privacy and tools.  
Du \etal \cite{du2020federated} discuss a comprehensive overview of the recent advances and applications of FL in the context of vehicular networks and IoT.
Chen \etal\cite{chen2021distributed} present a comprehensive survey on the deployment of distributed learning systems over wireless networks with a major focus on communication efficiency of all the different paradigms of distributed learning. In addition, 
Gafni \etal\cite{gafni2021federated} provide a detailed survey on the need for efficient signal processing to boost the performance of FL systems deployed at the edge. Hu \etal\cite{hu2021distributed} review the topic of distributed machine learning and its applications in wireless networks and communication efficiency.
Hellstrom \etal\cite{hellstrom2020wireless} present a survey that discusses the role of wireless communications for performance characterization of the FL.

It is evident from the aforementioned discussion and Table \ref{tab:1} that the majority of the survey articles focus either on applications of FL for wireless communications or on designing communication efficient FL protocols. The role of wireless communications for FL has been touched upon in relatively less number of survey articles e.g.,  in \cite{chen2021distributed} \cite{gafni2021federated} \cite{hellstrom2020wireless}.To the best of our knowledge, this is a first survey article that highlights the bidirectional relationship between FL and wireless communications by providing a holistic overview on the FL for wireless, wireless for FL and communication efficient FL. This bidirectional relationship is shown in Figure \ref{fig:1} that illustrates a scenario of the future wireless networks (e.g., a multiuser MIMO setup with multiple users that carry their local data sets) where a global model is trained in a distributed manner using FL. This global model is responsible for decision making and transforming the network into a self-sustaining network which manages and allocates resources and takes control decisions in real-time. However, the federated training would only be beneficial if efficiency of the wireless network is optimal as FL needs to transmit the model over uplink and downlink channels. \textit{We term this inter-dependency among FL and wireless communications as a bi-directional relationship. This paper emphasizes the fact that there needs to be frameworks and solutions in future which jointly optimize both FL and wireless communications considering the significance of this bidirectional relationship.}

Keeping aforementioned discussion in mind, our main contributions are summarized as follows:
\begin{enumerate}
\item[(i)]  An important  `bidirectional' relationship among FL and wireless networks has been explored in a comprehensive manner. None of the existing survey papers take into account this inter-dependency which is crucial in order to make them co-exist and achieve the goals of the next generation wireless communication systems.
\item[(ii)] We classify progress in this field into three categories: FL for wireless communications, Communication-efficient FL and Wireless communications for FL. This kind of fine-grained classification has been mostly overlooked and has not been discussed in a holistic manner.
\item[(iii)] For the first time, we discuss the role of the smart radio environments (i.e., IRS assisted environments) in the context of wireless for FL.

\item[(iv)] Our survey article also provides a discussion on the open problems and future directions of this interesting but less explored field, which we believe will help  researchers to put focus on the important but challenging problems.
\end{enumerate}

\section{Organization}
The organization of the rest of this survey article with its sections and subsections is shown in Figure 2 using a hierarchical diagram. This paper is mainly divided into nine sections. Section \ref{sec2} provides the necessary background knowledge required to understand the further sections of this paper. In particular, in this section, we provide theoretical details about the centralized and distributed learning strategies. In addition, we also present an elaborate theory of FL and its statistical perspective. Furthermore, Section 5 provides discussion about those techniques that can be used for communication-efficient FL. Section \ref{sec4} presents a detailed discussion on the influence of wireless communication on the performance of the FL. In Section 7 we discuss the impact, challenges and applications of FL for the advancements of wireless communication networks. Moreover,
Section 8 provides a discussion on all the open challenges and future directions which we believe are capable of driving breakthroughs in this domain. Finally, Section 9 concludes the survey article. 
\begin{figure}[t]
\centering
\includegraphics[width=0.4\textwidth]{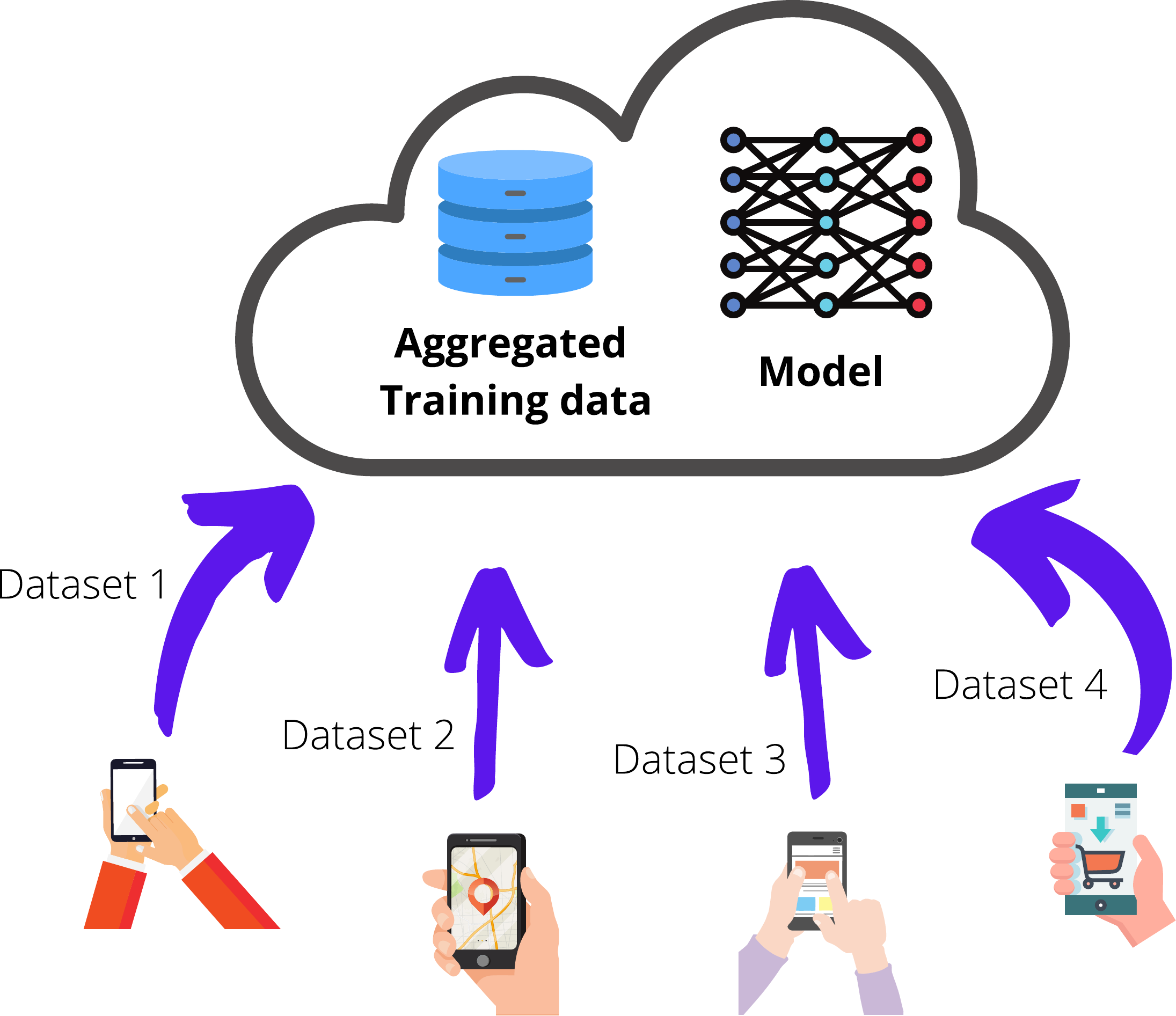}
\caption{Centralized training}
\label{fig:3}
\end{figure}
\begin{figure*}[!t]
\centering
\includegraphics[width=0.9\textwidth]{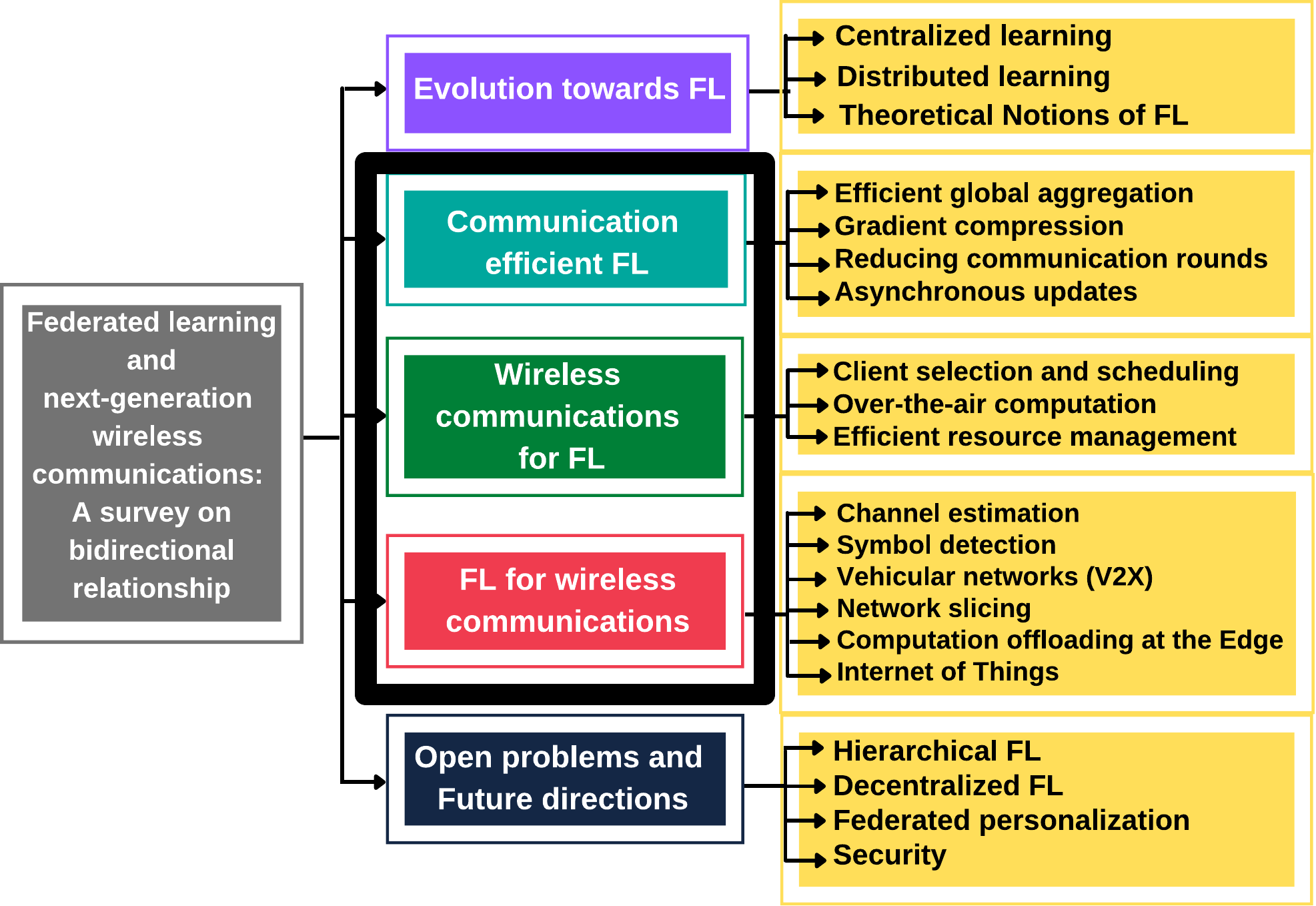}
\caption{Paper organization}
\label{fig:2}
\end{figure*}

\section{Evolution towards federated learning}\label{sec2} 
The exciting race for intelligence began almost 78 years ago, when threshold logic inspired by the human brain was used to create a computer model for neural networks. After that, the inventions of Back-propagation \cite{rumelhart1986learning}, Perceptron\cite{rosenblatt1958perceptron}, Hebbian learning \cite{hebb1949organization} and Hopfield Networks\cite{pribram2014hopfield} transformed the concept of neural networks into a promising field. However, due to the rise of shallow machine learning algorithms like Decision Trees and Support Vector machines, the progress in neural networks was on a stall for many years. Back in 2006, Hinton \etal\cite{hinton2006reducing} proposed the concept of Deep Belief Nets which was a stepping stone towards the deep learning revolution. Multiple advances, such as the ImageNet moment\cite{krizhevsky2012imagenet}, increased computing power in terms of powerful GPUs and the emergence of big data, have since moulded deep learning into one of the most successful fields in history.
In this section, we will go through a brief overview of the theoretical aspects of both centralized learning, distributed learning and federated learning.
\subsection{Centralized learning}
 In centralized training, the deep learning model is housed in a single PS and client edge devices send their entire data sets to the server. All of these client's local data sets are aggregated and loaded batch-wise to train the model as shown in Figure \ref{fig:3}. 

The most frequent way for training deep learning models is to employ Gradient Descent\cite{ruder2016overview}, a popular optimization algorithm that iterates over training data and gradually changes model parameters ($\theta$) in order to minimize a loss function ($L$). 

Following are variants of the Gradient Descent algorithm that are used widely in deep learning theory:
\begin{enumerate}
    \item \textbf{Vanilla gradient descent: }
        All of the samples from the training dataset are forward fed through the neural network in Vanilla / Batch gradient descent, thus there is just one parameter update step for the whole dataset. The learning rate $\eta$ determines the size of a parameter update step.
        \begin{equation}
            \theta = \theta - \Dot{\eta}{\nabla_\theta}{L(\theta)}
        \end{equation}
        It has a smooth optimization curve, but it would only be helpful in very small datasets because modern-day deep learning models contain millions of trainable parameters and training would take months with Vanilla gradient descent.
    \item \textbf{Stochastic gradient descent: }
        Only one random sample from the training dataset is forward transmitted and backpropagated upon in Stochastic gradient descent (SGD) to create one parameter update step. It has a fluctuating optimization curve due to its stochastic nature, but with an adequate learning rate, global minima may be attained in less time than Batch gradient descent.
        \begin{equation}
            \theta = \theta - \Dot{\eta}{\nabla_\theta}{L(\theta, x^{(i)}, y^{(i)})}
        \end{equation}
    \item \textbf{Mini-batch gradient descent: }
        In mini-batch gradient descent, the parameter update step works upon a random batch of training data of batch-size $n$. It is superior to Stochastic gradient descent because of its more steady convergence, and it is superior to Batch gradient descent because with an ideal batch size $n$, it can be used on bigger datasets.
        \begin{equation}
            \theta = \theta - \Dot{\eta}{\nabla_\theta}{L(\theta, x^{(i:i+n)}, y^{(i:i+n)})}
        \end{equation}
        
\end{enumerate}

\subsection{Distributed learning}
\begin{figure*}[t!]
\centering
\includegraphics[width=350pt]{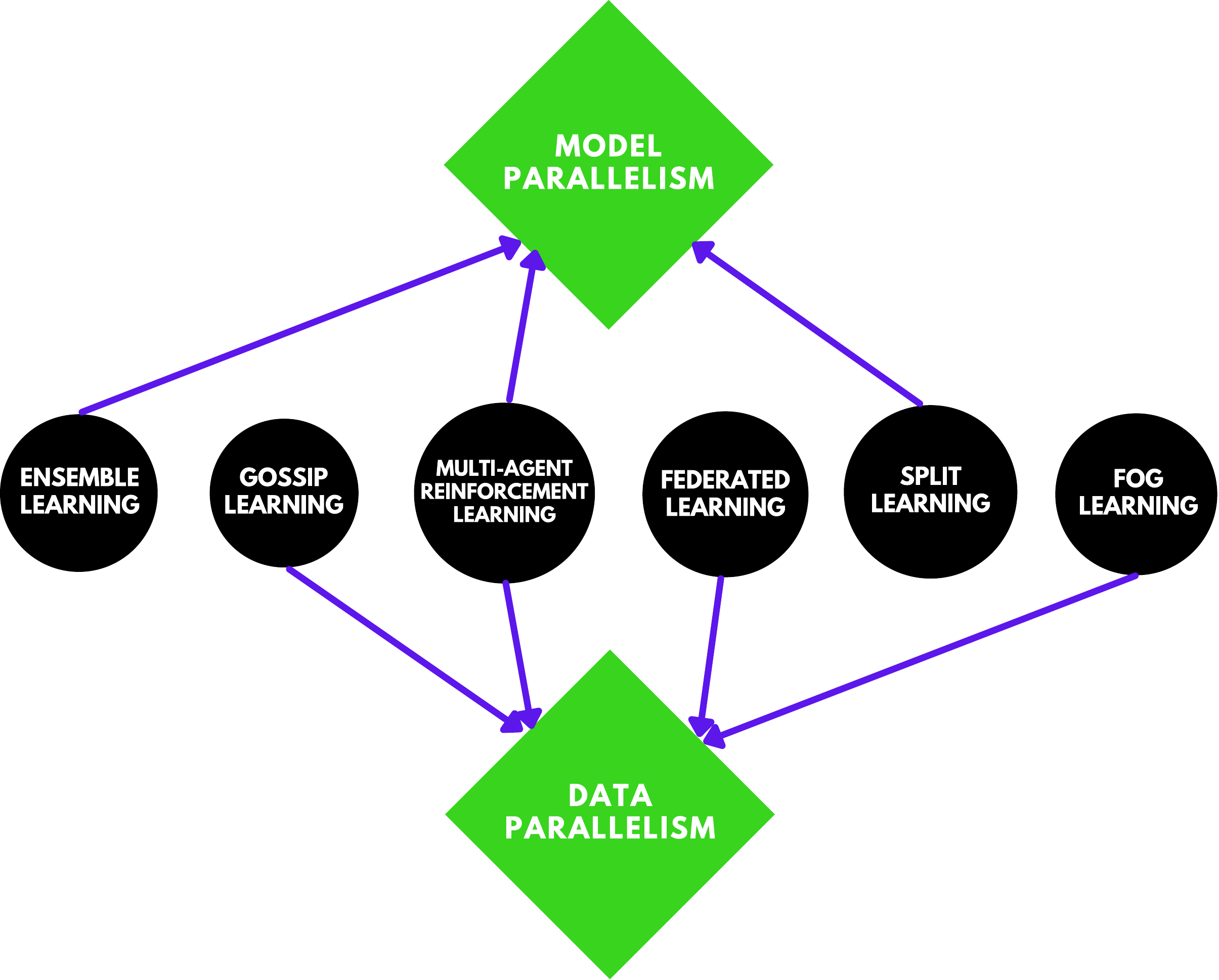}
\caption{Types of Distributed deep learning}
\label{fig:4}
\end{figure*}

As the datasets and models have been growing at an unprecedented rate, it is becoming impractical to train a model on a single machine\cite{l2017machine}. In order to overcome this challenge and to preserve privacy of the end users, distributed learning strategies are gaining popularity, as discussed earlier. Distributed learning is divided into two categories (as illustrated in Figure 4) dependent on which component of the training process is distributed: 
\begin{enumerate}
    \item \textbf{Model parallelism: } It refers to the technique of training large models by sharing the layers among different compute resources and training them using an efficient parallel pipeline.
    \item \textbf{Data parallelism: } It refers to splitting the large datasets into multiple parts and training the model(s) on those chunks of data in parallel.
\end{enumerate} 
Data parallelism is the most important paradigm as most of the wireless edge devices do not yet have the capability to utilize model parallelism due to computing constraints. 
Distributed learning is an important area, however it is still in its early stages of development. Most distributed training methodologies have been focused on using large data centers and dividing data and models within them, however this approach does not utilize the full advantage of distributed learning and is still computationally expensive. Moreover, this approach poses a threat to user's private data. 
The various types of distributed learning methods are illustrated in Figure \ref{fig:4} with categorization based on model or data parallelism.
\\
Ensemble learning is the process of creating and combining multiple models to address a specific machine learning problem. The obvious reason for the ensemble technique comes from human nature and our proclivity to collect and consider several viewpoints in order to make a complicated decision\cite{sagi2018ensemble}. As it uses multiple diverse models to train on the same dataset, it falls under pseudo model parallelism.
When environmental dynamics impact agent's decisions, they must learn about them and adapt their techniques based on their experiences with agent-to-environment and inter-agent interactions. Reinforcement Learning (RL) is important because of its exploration and exploitation capabilities. In RL, exploration assists agents in learning about the implications of their actions. Multi-agent RL is used to investigate the interactions of several agents in a comparable environment as they make decisions based on local observations. It can fall into the paradigm of data parallelism when each agent trains on decentralized datasets, and also fall under model parallelism when a large agent is parallelized over a compute instance for faster convergence.
FL is a distributed learning methodology that falls under the category of data parallelism \cite{mcmahan2017communication}. Each client has its own training dataset, which is not shared with the PS. As illustrated in Figure 2, a global model is delivered from the PS to all participating client devices and trained on each of their local datasets. As FL employs distributed data, it is a perfect method for training deep learning models over these private datasets of client devices. This would help to capture the local patterns of each user, making the model much more personalized and thereby making better decisions for the user's devices. For instance, the first real-world deployment of FL was observed in Gboard, the multilingual keyboard by Google presented in Android smartphones. FL made it possible to capture user behavior with local learning which eventually led to enhancing the future next-word predictions.
Split learning \cite{vepakomma2018split} is a type of model parallelism, where the neural network's layers are split and transferred to multiple devices to be trained locally. A deep neural network with a huge size cannot fit into the tiny memory of an edge device. By breaking a single model into numerous segments and spreading the lower portions among different clients holding raw data, Split learning addresses this problem. Each device uploads its model's cut-layer activations by linking the lower segments to a shared top segment maintained at a PS.
Gossip learning\cite{giaretta2019gossip} is a type of Distributed learning in which there is no requirement for a central entity to create a global model or to aggregate model updates from local learners. Similar to FL, it is a recent paradigm built as an on-device collaborative training architecture that does not need the transfer of raw data. 
Fog learning automatically spreads machine learning models to train throughout a network of nodes, from edge devices to cloud servers. Fog learning\cite{hosseinalipour2020federated} improves FL in three ways: on the network, in heterogeneity, and in proximity. It takes into account a multi-layer hybrid learning architecture made up of heterogeneous devices with varying degrees of proximity. The fog learning system automatically distributes machine learning model training over a continuum of nodes, from edge devices to cloud servers. Fog learning improves FL in three key ways: network, heterogeneity, and proximity. It analyses a multi-layer hybrid learning framework made up of heterogeneous devices located at different distances.
The parameters/gradients of the trained (Edge) models are uploaded to the server and aggregated with the Global model using specific aggregation methods. 
\subsection{Theoretical notions of Federated learning}
\subsubsection{FL training steps}
A typical FL training round is a multi-stage procedure orchestrated by a global PS. A general template\cite{kairouz2019advances} for any FL round involves the following steps:
\begin{enumerate}
    \item \textbf{Client selection: }The PS selects a set of clients based on some criteria.
    \item \textbf{Global model broadcasting: }The participating client devices download the global model present at the PS.
    \item \textbf{Local training: }The clients perform on-device computation using any form of gradient descent and update their respective local models.
    \item \textbf{Local updates transfer: }After each FL round on a client device, the parameters / gradients of the locally trained model are transmitted to the PS.
    \item \textbf{Global aggregation: } The PS collects parameters / gradients from multiple clients and aggregates them to update the global model.
\end{enumerate}

\subsubsection{Statistical perspective}
\begin{figure*}[t!]
\centering
\includegraphics[width=350pt]{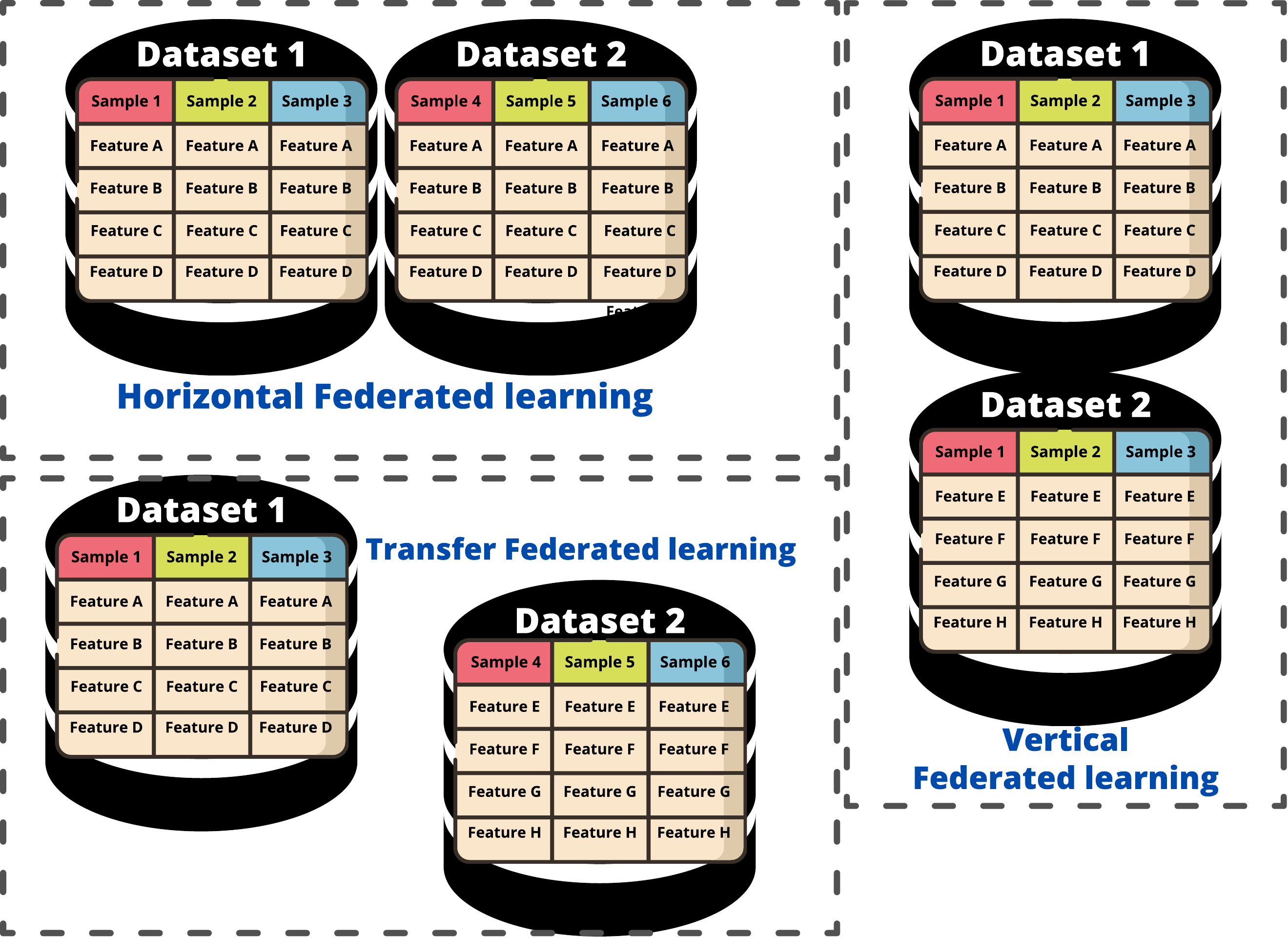}
\caption{Types of Federated learning}
\label{fig:5}
\end{figure*}

Federated training occurs under several constraints such as non-independent and identical distributed (non-IID), unbalanced and massively distributed local datasets of client devices along with imperfections of the wireless network.
FL has been classified into three categories based on statistical aspects of the accessible decentralized data, as illustrated in Figure \ref{fig:5}. 
\begin{enumerate}
    \item \textbf{Horizontal FL: }This is the scenario when the decentralized datasets have the same feature space, but differ in sample space. A typical example would be the use-case of FedFace\cite{aggarwal2021fedface}, in which a global optimal face recognition model gets trained by using private facial data of multiple client devices each employ the same facial application.
    \item \textbf{Vertical FL: }It refers to the scenario when the decentralized datasets have same sample space, but they differ in feature space. An example of Vertical FL would be an e-commerce company and a social media platform trying to collaboratively train a model to predict customer's purchase behaviour. Here, the two companies have a large number of common users which make them have the same sample space, but they have non-overlapping features as an e-commerce company would generally track a customer's product viewing and purchase history whereas the social media app would track following, likes and social behaviour of the customer.
    \item \textbf{Transfer FL: }It refers to the case when the decentralized datasets differ from each other in terms of both sample and feature space. An example of Transfer FL is a recommendation model that learns from various app-usage behaviours of two users in different nations. Since they use different applications on their devices, resulting in a distinct decentralized datasets without any overlap of sample and feature space among them.
\end{enumerate}
\section{Communication efficient Federated learning}\label{sec3}
One of the main goals for designing FL was to reduce communication overhead, however deployment of FL in practice still requires considerable communication resources. Hence, in the sequel,  we discuss the major advances and efforts that are made in the literature to make FL more communication efficient and deployable.
\subsection{Efficient global aggregation}
The technique used for globally aggregating parameters/gradients from all the clients needs to be statistically efficient in order to reduce the time required for convergence of the global loss during FL training.
In literature, several global model aggregation and training techniques have been proposed, with Federated Averaging (FedAvg) being the standard.
In FedAvg\cite{mcmahan2017communication}, at first the PS initializes the global model and selects a random set of $M$ clients out of all of the available client devices. A global model with random weights is initialized at the PS and passed on to each of the client devices. At each participating device, the local datasets are divided into mini-batches each having size of $B$ and the local training phase starts in parallel across all the participating clients. In each epoch of the local training, each mini-batch is passed on a step of gradient descent for minimizing the local loss function. All the local weights from each client are transferred to the PS via \textit{wireless links} and the PS aggregates these weights/gradients together by averaging them.

FedProx \cite{li2018federated} is an improved version of the FedAvg algorithm by tolerating partial work and varying resource constraints among the participating client devices. In other words, not all devices have the computing power to handle model training at the same scale, therefore FedProx includes a variable called $\gamma$-inexactness that represents how much work the device can manage. It also includes a proximal factor, which helps to address statistical heterogeneity by limiting local updates to be closer to the global model without needing an explicit definition of the number of local epochs. It also allows for the safe integration of varying amounts of local effort induced by system heterogeneity.

Federated Matched Averaging (FedMA) \cite{wang2020federated} is a novel statistical layers-wise FL approach for current Convolutional Neural Networks (CNNs) and Long short term memory networks (LSTMs) that use Bayesian non-parametric approaches to adjust the data heterogeneity. It considers permutation invariance of each neuron in the model before the global aggregation to achieve globally adaptive model size. 

FedPAQ\cite{reisizadeh2020fedpaq} is a communication-efficient FL approach that incorporates Periodic Averaging and Quantization. It involves periodic averaging, in which models are updated locally at devices and only periodically averaged at the PS. It takes into account partial device involvement, in which only a small percentage of devices engage in each round of training. Moreover, it utilizes quantized message-passing where the edge devices quantize their updates before uploading to the PS.

\subsection{Gradient compression} 
FL requires high network bandwidth during the gradient transfer phase over the uplink wireless channels and model transfer phase from PS to client devices over downlink channels. Widely used models with deep architectures such as transformers have upto a billion parameters, which would lead to a massive communication bottleneck if used for standard FL training. Gradient compression is a promising solution which involves selective transmission of gradients during the FL training rounds in order to reduce the effect of the communication bottleneck.
Deep Gradient Compression (DGC) was presented by Lin \etal\cite{lin2017deep} that leveraged several advanced techniques and achieved  up to 600 times compression of the gradients without significant degradation in performance. The authors used gradient sparsification to send only those gradient updates in each communication round which are larger than a threshold to ensure that most important gradients are transmitted. The total communication cost in distributed training depends on the number of bits in each transmission and number of communication rounds. To reduce the number of bits in each round, Sattler \etal\cite{sattler2019sparse} proposed sparse binary compression (SBC) which involved a novel binarization method for gradient sparsification and was able to achieve up to 37208 times reduction in bits with only 1\% reduction in model accuracy. Furthermore, Cui \etal\cite{cui2020creat} used K-means clustering for selectively uploading only a small fraction of gradients considering the fact that only few gradients are distant from `zero' value. The authors quantized these selected gradients and verified them using blockchain based method to ensure no forged gradients are uploaded to the PS. Tang \etal\cite{tang2019doublesqueeze} proposed DoubleSqueeze, a bidirectional (from the clients to the PS and vice-versa) compression technique for distributed learning in an IID setting. The approach consists of transmitting a gradient compression coupled with error compensation for both uplink and downlink channels. Artemis\cite{philippenko2020bidirectional}, a theoretical framework for bidirectional compression was proposed by Philippenko \etal in order to be robust under Non-IID data and partial client participation. Abrahamyan \etal\cite{9451554} designed an autoencoder with a lightweight architecture which captures the common patterns in the gradients of the different distributed clients and achieved a 8095 times compression which is 8 times more than DGC. Entropy based gradient compression scheme was proposed by Kuang \etal\cite{kuang2019entropy} which consisted of an entropy based threshold selection method and a learning rate correction algorithm. Entropy is a well known metric from information theory which here measures the uncertainty or disorder of the gradients. Generally, the entropy of a layer's gradient is low if the gradient includes lesser information. Using the obtained entropy information and QuickSelect algorithm, the threshold is calculated and only those gradients with absolute value above the threshold are transmitted in that communication round. The results in \cite{kuang2019entropy} showed that up to 1000 times gradient compression is achievable while keeping the accuracy of the model nearly unchanged. Fast FL was proposed by Nori \etal\cite{nori2021fast} which attempts to jointly consider the local weight updates and gradient compression tradeoff in FL. They formulate their problem as an optimization objective of minimizing learning rate error under the constraints of computation and communication interdependency in FL.

\subsection{Reducing communication rounds}
It is well-understood that reduction in the number of communication rounds results in improving communication-efficiency. In the literature, several attempts have been made to make the FL communication efficient. Luping \etal \cite{luping2019cmfl} presented Communication Mitigated FL (CMFL) that identifies the importance of each communication round and prevents the less important model updates to be transferred over the uplink while having a  guaranteed convergence. CMFL uses the most recent global model and compares it with the present local model update. If the number of same sign parameters is high, then intuitively it ensures that the communication round is of high relevance and is not an outlier with respect to the global updates.
Jiang \etal\cite{jiang2020adaptive} presented Adaptive periodic averaging in which they showed that by communicating more in the earlier rounds of FL and gradually reducing the number of communication rounds at the later stages of the training gives a much better convergence due to the fact that variance of model parameters is larger during the initial rounds and becomes much lower during the later stages. Furthermore, Yao \etal\cite{yao2018two} designed a method termed two-stream FL where they introduced maximum mean discrepancy (MMD) loss in order to reduce the number of communication rounds of FL. This novel MMD loss function measures the distance between the outputs of the received global model and trained local model in order to force the local model to gather more knowledge from the local data available for faster convergence, thus resulting in lesser number of communication rounds. Feature fusion FL was developed recently\cite{8803001} in which features from global and local models were aggregated in order to achieve reduction in number of communication rounds by 60\% while maintaining equivalent accuracy. In this approach, a local image serves as input to the local feature extractor as well as a global feature extractor, and a fusion operator translates those extracted global and local features into a fusion feature space and each feature fusion module is sent to the PS for global aggregation.
\subsection{Asynchronous updates}
In the classical FL setting as introduced in \cite{mcmahan2017communication}, the communication type is synchronous where the global model is sent to all clients and then the training process starts, which leads to a large communication delay due to massive number of participating devices. All devices are not capable enough at all times to support federated training due to factors such as power discharge or network failure. Hence, a training process of the FL can stall that may ultimately lead to poor communication efficiency. Therefore, development of FL communication rounds in an asynchronous way is of utmost importance. Recently, some progress has been made in literature on asynchronous FL.
A Semi-Asynchronous Federated Averaging (SAFA)\cite{wu2020safa} protocol was developed in which prior to every local training round, the PS classifies all available clients into three categories: (i) Up-to-date clients who have completed previous training rounds successfully, (ii) Deprecated clients who have outdated models in comparison with the latest global model, and (iii) Tolerable clients whose local model version is outdated but not too old. SAFA only allows up-to-date and deprecated clients to synchronize with the PS while tolerable clients are trained in an asynchronous manner. Here, deprecated clients are forced to synchronize in order to make sure that their outdated local updates do not poison and affect the latest global model. Chen \etal\cite{chen2019communication} designed an asynchronous FL technique in which layers of the neural networks are classified into deep and shallow layers. In this proposed protocol, more often, the local updates constitute weights from the shallow layers as compared to the deeper layers because the deeper ones learn dataset specific characteristics whereas shallow ones learn generalized patterns. The authors also used a temporally weighted aggregation approach where the recent updates are considered of more importance and thus assigned higher weights.

\section{Wireless communications for Federated learning}\label{sec4}
As aforementioned discussion and illustration given in Figure 1, wireless communication is critical for FL, however it poses a difficulty for model training since channel randomness skews each client's model update, and numerous client updates cause severe interference due to restricted bandwidth. In this section, we discuss techniques of wireless communications that can be designed  to improve performance of the FL.
\subsection{Client selection and scheduling}
As discussed in previous section that the standard FL\cite{mcmahan2017communication} training process can become highly inefficient when resource constraints of real-world wireless networks are taken into account. Edge devices have limited and varying compute resources and are connected to wireless networks with dynamically changing conditions due to fading and/or shadowing. Under such limitations, using all devices for FL is not reasonable and thus to tackle this issue, client selection methodologies have been proposed by several researchers. In \cite{nishio2019client}, a new FL protocol is proposed, termed as `FedCS' that uses a two step client selection procedure by requesting the clients their edge device's resource information in real-time and a subsequent client selection step based on the received information. The client selection step has been framed as a multi-objective maximization problem for accepting as many relevant client updates as possible.  This maximization problem is solved by a Heuristic greedy algorithm.
Furthermore, `FedMCCS' is proposed in \cite{abdulrahman2020fedmccs} that works as a multi-criteria based client selection procedure in order to predict if a client is capable of participating in the federated training, taking into account the client device's resource information such as CPU, free memory, time and energy. The authors show experimentally that FedMCCS is able to reduce the number of communication rounds, maximize the number of clients, optimizes the network traffic as tries to minimize the number of discarded rounds.
Among the large number of client edge devices participating in FL, there is a high possibility of unreliable updates be it intentionally for poisoning attacks or unintentionally due to resource constraints. Keeping this in mind, Kang \etal\cite{kang2020reliable} discussed the idea of selecting clients based upon a new metric termed as reputation using a consortium blockchain based scheme. After broadcasting the global model, the PS receives the identity and resource information from the clients and calculates the reputation metric. These reputation values are fetched either from the historical logs of client participation from the PS, or from an open access consortium blockchain of multiple PSs. These values depend on interaction frequency, timelines and effects of each client.
Moreover, three standard scheduling algorithms (Round-robin, proportional fair and random scheduling) were compared for efficiency in terms of client scheduling for FL in a study by Yang \etal\cite{yang2019scheduling}. In this work, convergence of FL is considered in a large-scale wireless setting with practical network limitations such as interference. More specifically, in random scheduling (RS), the PS selects a subset of clients at random in each update transmission round. In round-robin scheduling (RR), the PS clusters all the available clients into groups in each communication round and assigns them to utilize the radio channels to communicate their respective updates. In proportional-fair scheduling (PF), the PS selects a subset of clients using a policy that ranks them based on signal-to-noise ratio (SINR) values. It is demonstrated through numerical and experimental simulations that PF scheduling outperforms the other two in terms of iteration time when the SINR is high, whereas RS performs best under low SINR conditions. 
Age-based scheduling algorithms have also been investigated \cite{yang2020age,buyukates2020timely} which select and transmit only those client updates which are less stale (i.e, more recent).
All the above discussed client selection and scheduling methods have been summarized in Table \ref{tab:3}.

\begin{table*}[t]
\caption{Client selection and scheduling algorithms in literature} \label{tab:3}
    \centering
    \begin{tabular}{| m{3cm} | m{5cm} | m{4cm} | m{4cm} |}
        \hline
        \textbf{Paper(s)} & \textbf{System model} & \textbf{Metric} & \textbf{Results / Objective}\\
        \hline
        FedCS\cite{nishio2019client}
        & 
         \begin{itemize} 
            \item Channel information
            \item Non IID data
            \item Resource constraints
        \end{itemize}& 
        \begin{itemize}
            \item Accuracy
            \item Elapsed time
            \item Number of clients
        \end{itemize}& 
        \begin{itemize} 
            \item Client maximization
            \item Time minimization
        \end{itemize} \\
        \hline
        FedMCCS\cite{abdulrahman2020fedmccs}
        & 
         \begin{itemize} 
            \item Location based sampling
            \item Non IID data
            \item Device information
            \item Resource constraints
        \end{itemize}& 
        \begin{itemize}
            \item Resource utilization
            \item Accuracy
            \item Number of clients
            \item Elapsed time
        \end{itemize}& 
        \begin{itemize} 
            \item Client maximization
            \item Number of rounds minimization
            \item Discarded round minimization
            \item Network traffic optimization
        \end{itemize} \\
        \hline
        Kang \etal\cite{kang2020reliable} & 
         \begin{itemize} 
            \item Client reliability
            \item Poisoning attacks
            \item Real-time client monitoring
        \end{itemize}& 
        \begin{itemize}
            \item Reputation
        \end{itemize}& 
        \begin{itemize} 
            \item Secure FL
            \item Client maximization
        \end{itemize} \\
        \hline
        Yang \etal\cite{yang2019scheduling}
        & 
         \begin{itemize} 
            \item RS, RR and PF scheduling
            \item Channel information
            \item Resource constraints
        \end{itemize}& 
        \begin{itemize}
            \item Number of rounds minimization
            \item Elapsed time
            \item Accuracy
        \end{itemize}& 
        \begin{itemize} 
            \item Comparative study
        \end{itemize} \\
        \hline
        Yang \etal\cite{yang2020age}
        
        Buyukates \etal\cite{buyukates2020timely}
        & 
         \begin{itemize} 
            \item Staleness 
            \item Channel information
            \item Power constraint
        \end{itemize}& 
        \begin{itemize}
            \item Age of update
        \end{itemize}& 
        \begin{itemize} 
            \item Low complexity
            \item Client maximization
        \end{itemize} \\
        \hline
        
    \end{tabular}
\end{table*}
\begin{figure*}[!t]
\centering
\includegraphics[width=0.7\textwidth]{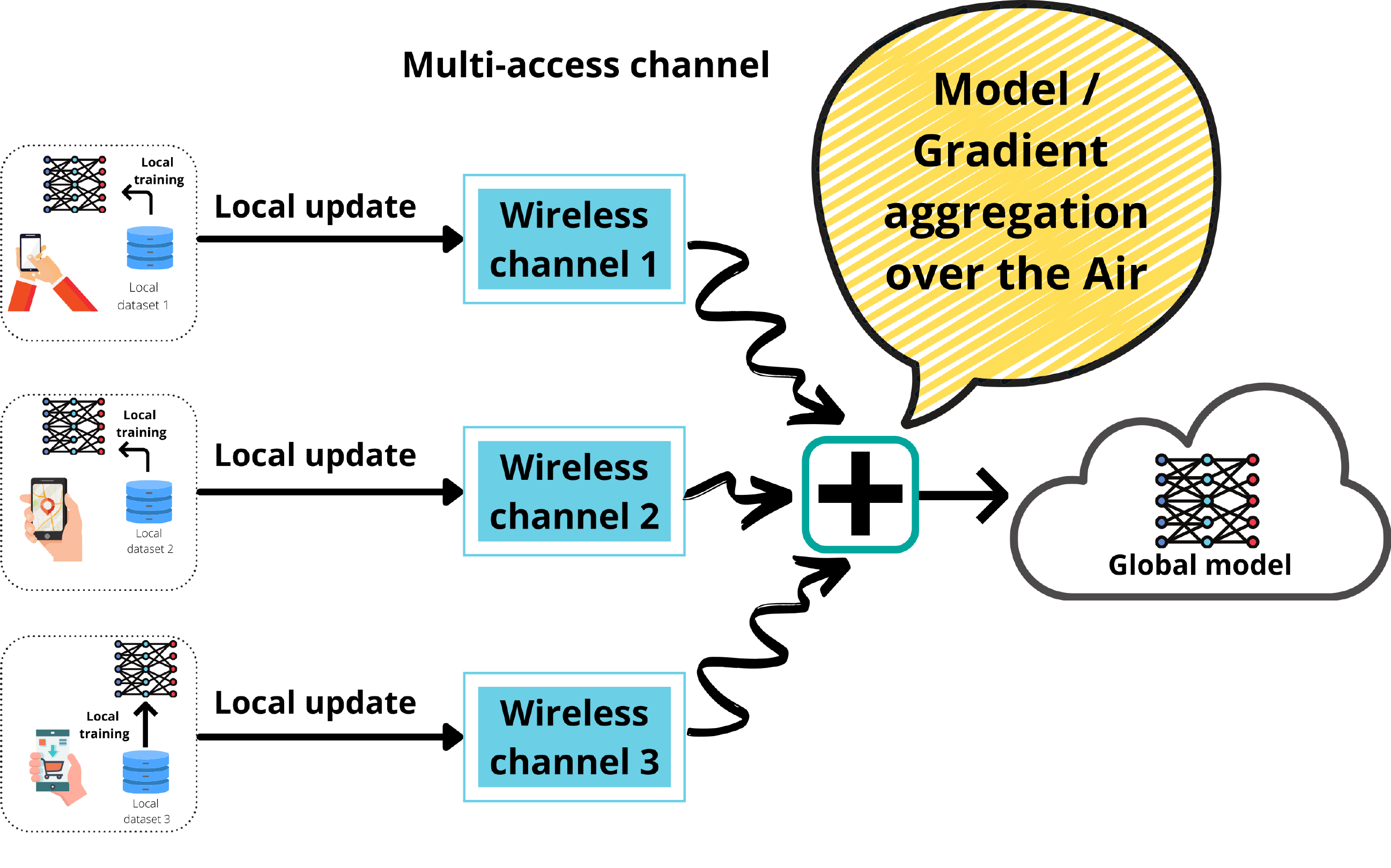}
\caption{AirComp based FL mechanism}
\label{fig:6}
\end{figure*}
\subsection{AirComp} \label{aircomp}

A wireless channel has several hidden properties that are not fully explored in the existing literature. Nevertheless, these properties can be leveraged to make edge machine learning much more communication efficient.
Distributed computation over multi-access wireless channels has been explored in wireless communications\cite{nazer2007computation, goldenbaum2013harnessing}. It is based on utilizing the phenomena of analogue signal superposition across wireless channels for computing nomographic functions over the air (AirComp), as illustrated in Figure \ref{fig:6}. Keeping in mind the fact that the model/gradient aggregation operation is a nomographic function\cite{hua2019device}, this AirComp idea has been investigated in literature for use in the gradient aggregation phase of FL. Analog gradient aggregation schemes provide a huge improvement in terms of communication latency compared to standard orthogonal frequency division multiple access (OFDMA) based schemes\cite{zhu2019broadband}.
\begin{table*}[!htbp]
\caption{FL with AirComp methods in literature}
    \centering
    \begin{tabular}{| m{3cm} | m{5cm} | m{4cm} | m{4cm} |}
        \hline
        \textbf{Paper(s)} & \textbf{System model} & \textbf{Metric} & \textbf{Results / Objective}\\
        \hline
        Zhu \etal\cite{zhu2019broadband}
        & 
         \begin{itemize} 
            \item Channel Noise and fading
            \item Large number of clients
            \item OFDM modulation
            \item Amplitude alignment
        \end{itemize}& 
        \begin{itemize}
            \item Recieve SNR
            \item Truncation ratio
            \item Latency reduction ratio
            \item Reliability ratio
        \end{itemize}& 
        \begin{itemize} 
            \item Convergence gurantee
            \item Latency minimization
            \item Path loss minimization
        \end{itemize} \\
        \hline
        COTAF\cite{sery2020cotaf} & 
         \begin{itemize} 
            \item Noisy channels
            \item Pre-coding and scaling
        \end{itemize}& 
        \begin{itemize}
            \item Convergence rate
            \item Model loss
        \end{itemize}& 
        \begin{itemize} 
            \item Convergence gurantee
            \item Channel noise mitigation
        \end{itemize} \\
        \hline
        Amiri \etal\cite{amiri2020machine}
        & 
         \begin{itemize} 
            \item Power allocation
            \item Error monitoring
            \item Non IID data
            \item Resource constraints
            \item Gradient sparsification
        \end{itemize}& 
        \begin{itemize}
            \item Accuracy
        \end{itemize}& 
        \begin{itemize} 
            \item Bandwidth efficient
            \item Latency minimization
        \end{itemize} \\
        \hline
        Yang \etal\cite{yang2020federated}
        & 
         \begin{itemize} 
            \item Ideal channel assumption
            \item Non IID data
            \item Resource constraints
        \end{itemize}& 
        \begin{itemize}
            \item Feasibility
            \item Accuracy
            \item Number of clients
        \end{itemize}& 
        \begin{itemize} 
            \item Client maximization
            \item Feasibility maximization
        \end{itemize} \\
        \hline
        Vu \etal \cite{vu2020cell}
        & 
         \begin{itemize} 
            \item Cell-free MIMO
            \item Joint resource, accuracy and training time optimization
        \end{itemize}& 
        \begin{itemize}
            \item Training time
            \item Accuracy
            \item Data rate
        \end{itemize}& 
        \begin{itemize} 
            \item Training time minimization
        \end{itemize} \\
        \hline
        Wang \etal\cite{wang2021federated} & 
         \begin{itemize} 
            \item IRS assisted AirComp
            \item Two-step low rank optimization
        \end{itemize}& 
        \begin{itemize}
            \item Client frequency
            \item Mean Squared error
        \end{itemize}& 
        \begin{itemize} 
            \item Client maximization
            \item Faster convergence
        \end{itemize} \\
        \hline
        Yu \etal \cite{yu2020optimizing} & 
         \begin{itemize} 
            \item IRS assisted AirComp
            \item Large-scale C-RAN
            \item Joint optimization of reflection phase and linear detection vector
        \end{itemize}& 
        \begin{itemize}
            \item Mean squared error
        \end{itemize}& 
        \begin{itemize} 
            \item MSE minimization
        \end{itemize} \\
        \hline
    \end{tabular}
 \label{tab:4}
\end{table*}
Convergent over-the-air FL (COTAF) algorithm is proposed in \cite{sery2020cotaf} and is shown to converge successfully even over the shared noisy wireless channels. COTAF employs temporal precoding and scaling methods to reduce the impact of channel noise while transmitting local updates to the PS over the uplink channels. 
Amiri \etal\cite{amiri2020machine} presented Analog distributed stochastic gradient descent (ADSGD) which sparsifies local model gradients and encodes them to a low dimensional space to be efficient for over-the-air computation under limited bandwidth of the wireless channel. They formulate DSGD as a distributed wireless computation problem for addressing power and bandwidth constraints while considering both physical layer properties as well as the model's convergence. The authors showed that without loss in performance, ADGSD is much more bandwidth efficient as compared to the other digital schemes.

Moreover, Broadband Analog Aggregation (BAA) framework was introduced by Zhu \etal\cite{zhu2019broadband} which utilized the air computation principle
over broadband multi-access wireless channels. BAA was designed with an aim to reduce communication latency. BAA considers the scenario of a single cell network where each device transmits a large local model update over a broadband channel in blocks by linear modulation. Furthermore, in order to deal with fading and interference, BAA also utilizes orthogonal frequency division multiplexing (OFDM) to split the channel into sub-channels. In order to make sure identical amplitudes of local model updates are transmitted, BAA uses a broadband channel inversion mechanism for reliable analog transmission.

A joint beamforming and client selection optimization approach was used by Yang \etal\cite{yang2020federated} to maximize the number of client devices satisfying the thresholds of satisfactory performance in AirComp based FL training. The corresponding non-convex combinatorial optimization problem is transformed into a sparse and low rank problem using a sparse representation for the objective function and matrix lifting technique. A difference of convex functions based algorithm was proposed by the authors which proved to be better than conventional approaches.
However, an important aspect often overlooked by the above methods is that gradient statistics differ across training iterations and feature dimensions and are unknown in advance. Zhang \etal \cite{zhang2021gradient} investigated the power control problem for over-the-air FL by taking into consideration the effects of gradient statistics. Their goal was to reduce aggregation error by maximizing transmit power at each device while keeping average power limits in mind. When we have gradient statistics, we can find the optimal policy in closed form. Notably, they demonstrate that the optimal transmit power is continuous and decreases monotonically with the squared multivariate coefficient of variation of gradient vectors.

Cell-free massive MIMO is a novel architecture for multi-user MIMO that is recently proposed\cite{zhang2019cell}. When using cell-free massive MIMO, base stations are deployed to serve users in a cooperative manner using the same time and frequency resources. The backhaul network connects all access points making it possible to synchronize conjugate beamforming in downlink and matching filtering in uplink. The primary benefit of using a cell-free design is to provide coverage to the large area because of the large number of base stations. To support any FL architecture, Vu \etal \cite{vu2020cell} presented a new approach for cell-free massive MIMO networks. This plan enables for each iteration of the FL framework to occur in a large-scale coherence time to provide a stable FL process.

IRS is a planar surface made up of low-cost, passive reflecting elements each of which can give rise to an amplitude or phase change to the incident signal and collectively these changes result in 3-D reflect beamforming\cite{wu2019towards}. IRS helps to mitigate the effect of wireless channel fading which in turn makes the network energy efficient and improves the signal strength. In a study\cite{waqarperformance}, it was demonstrated that an IRS-assisted communication system outperforms other baseline systems even with a small number of discrete phase shifts. IRS is capable of solving the problem of AirComp systems suffering a lot from challenging signal propagation conditions\cite{jiang2019over}. AirComp has been proven to be a promising method for enhancing FL performance under tough constraints of real-world wireless communication systems and improving AirComp with IRS would result in much more increase in FL performance. Wang \etal\cite{wang2021federated} showed IRS based AirComp-FL was able to achieve a much lower training loss and higher prediction accuracy than conventional baseline methods. Their method's objective was to simultaneously optimize device selection, IRS's phase shifts and the aggregation beamformer so as to cancel out the higher model aggregation errors when the number of client devices are maximized and the optimization problem was solved using a difference of convex (DC) algorithm. Furthermore, Yu \etal \cite{yu2020optimizing} discussed the benefits of IRS assisted AirComp in a cloud radio access network. The system model of \cite{yu2020optimizing} consisted of distributed access points (APs) to which the local model updates were sent and each AP forwarded the received update signals to the PS through the finite capacity fronthaul link. The authors designed an iterative algorithm for optimizing the reflecting phases of the IRS along with linear detection vector of the global PS. The aforementioned AirComp techniques for the FL have been summarized in Table. \ref{tab:3}.

\vspace{10 pt}
\subsection{Efficient resource management}
Client devices have many resource constraints, among  which battery depletion is one of the most fundamental challenges for the success of FL. This opens a research problem of energy optimization for FL which has been discussed by a few research articles. Zhan \etal\cite{zhan2020experience} formulated it as optimization of the weighted sum of the energy consumed and the time taken for each round of local training. Even IID local datasets of approximately same size can take different amounts of time for the local training due to varying amount of processing power and network quality available at each client. In a synchronous FL setup, the devices which take lesser time to update would have an unnecessary idle time to communicate the update. To solve the joint optimization problem, the authors used an actor-critic based deep reinforcement learning (DRL) algorithm instead of the conventional heuristic algorithms due to the unpredictable and highly dynamic conditions of the real-world wireless networks. The DRL agent selects an action using the learnt policy based upon the present states where state space includes a set of historical bandwidth information. The reward function resembles negative of the system cost so as to guide the DRL agent towards minimizing the system cost. The experimental results show a 40\% reduction in the system cost compared to previous state-of-the-art approaches. Furthermore, a novel resource allocation framework termed `resource rationing' was introduced by Shen \etal\cite{shen2021resource} in which it was highlighted that each learning round has a different importance level towards the final performance of the system due to the fact that factors like bandwidth, number of clients and energy limits are unique in each client device. Resource rationing is built upon the ``later-is-better" principle indicating that there is a significant performance boost if resources are reserved at the early stages of training and then utilized as much as possible in the later rounds. Most of the works in literature have considered a system model assuming only one FL service, but once FL systems start getting deployed at scale, there would be multiple simultaneous FL services co-existing within the same wireless network. Allocating resources in this multi-service FL scenario is a tough challenge due to sharing of the same spectrum among those services. In a study\cite{xu2021bandwidth}, it has been highlighted that multi-service FL's efficiency depends upon both intra-service and inter-service spectrum allocation. Intra-service refers to the general client-level resource allocation which we had been discussing prior to this. Inter-service refers to the wireless spectrum allocation among multiple active FL services where there can be two scenarios. Under the first scenario, the FL service providers are assumed to be cooperative and the wireless network operator fairly distributes the bandwidth which in turn resembles the same optimization problem as of the intra-service case with the FL service providers acting as users and has been solved using a distributed algorithm designed by the authors. The second scenario considers selfish FL service providers who misreport their resource constraints and workload to get a higher share of the spectrum to boost its own performance, which may lead to performance degradation among the other FL services sharing the spectrum. To solve this issue, the authors designed a multi-bidding auction mechanism and experimental results show that their algorithm's performance exceeds previous benchmarks. The summary of the aforementioned efficient resource management techniques for the FL is given in Table. \ref{tab:4}.
\begin{table*}[t]
\caption{Resource management techniques in literature}
    \centering
    \begin{tabular}{| m{2cm} | m{5cm} | m{5cm} | m{4cm} |}
        \hline
        \textbf{Paper(s)} & \textbf{System model} & \textbf{Metric} & \textbf{Results / Objective}\\
        \hline
        Zhan \etal\cite{zhan2020experience}
        & 
         \begin{itemize} 
            \item Dynamic network environment
            \item Client-side resource management
            \item Synchronous FL
        \end{itemize}& 
        \begin{itemize}
            \item Average computational energy
            \item Average training time
            \item Client CPU-cycle frequency
            \item Average system cost
        \end{itemize}& 
        \begin{itemize} 
            \item Energy efficiency
            \item Faster convergence
        \end{itemize} \\
        \hline
        Shen \etal\cite{shen2021resource} & 
         \begin{itemize} 
            \item "Later is better" principle
            \item Generalized system model
        \end{itemize}& 
        \begin{itemize}
            \item Bandwidth
            \item Client frequency
            \item Accuracy
        \end{itemize}& 
        \begin{itemize} 
            \item Client maximization
        \end{itemize} \\
        \hline
        Xu \etal\cite{xu2021bandwidth}
        & 
         \begin{itemize} 
            \item Multi-service FL scenario
            \item Two-level allocation
            \item Selfish service providers
            \item Multi-bid auction
        \end{itemize}& 
        \begin{itemize}
            \item Training time
            \item Bandwidth ratio
            \item FL frequency
        \end{itemize}& 
        \begin{itemize} 
            \item Number of rounds minimization
            \item Fair bandwidth allocation
        \end{itemize} \\
        \hline
        
    \end{tabular}
 \label{tab:5}
\end{table*}
\section{Federated learning for Wireless communications}\label{sec5}
In this section, we will discuss the major applications of FL for improving the future wireless communication networks. 
\subsection{Channel estimation}
Channel estimation plays a critical role in optimizing link performance in wireless communication systems. With the advent of mmWave in 6G, acquiring the channel state information becomes a challenging problem due to massive number of antennas, high bandwidth and complex design of transceiver \cite{hassan2020channel}. In the earlier generation of wireless communication systems, channels are estimated through machine learning based shallow algorithms by formulating a regression problem received pilot signals as the input data and the channel state information (CSI) as the output data. In addition to this, support vector machine (SVM) algorithms have been used widely in literature for channel estimation\cite{sanchez2004svm, charrada2012complex, 5159013, garcia2006support, 9148630} due to their ability to handle non-linear relationships between input and output data. Due to the increased complexity and non-linearity in current 5G and future 6G wireless communication channels, deep learning evolved as a potential solution for channel estimation. 
Most of the proposed deep learning models for channel estimation use centralized architectures, which can generate a huge overhead on the network, thus creating the need for decentralized learning. Elbir \etal\cite{elbir2020federated} showed that models trained with FL for channel estimation have 16 times lesser network overhead as compared to the models trained with centralized learning. The authors trained a convolutional neural network (CNN) based model for estimating the channel matrix in both standard and IRS assisted MIMO settings. 

\subsection{Symbol detection}
Symbol detection refers to the mapping of symbols received through a wireless channel to decoded clean symbols.  Centralized learning based symbol detection has been explored widely in literature due to its efficiency in end-to-end learning and detecting symbols without the need of channel information which is otherwise required in conventional algorithmic approaches. However,  centralized learning suffers from high network overload due to the huge size of the received symbol data. FedRec, a downlink fading symbol detector, based on decentralized data was proposed by Mashhadi \etal \cite{mashhadi2020fedrec}. It consists of neural networks under a fading channel system based on the maximum a posteriori probability (MAP) rule. It utilizes as a collaborative training technique that utilises channel diversity among several users through FL.

\subsection{Vehicular networks}
In complex Quality of Service (QoS) real-world settings, emerging vehicular networks incorporate a huge volume of vehicle sensor data and applications. Till date, such systems only considered centralized learning which is not a scalable and efficient solution due to privacy and network overload concerns in resource constrained vehicular networks\cite{du2020federated}. FL is ideal for these requirements as it can efficiently utilize decentralized data and compute from different vehicles in a privacy preserving manner. Hence, Mashhadi \etal\cite{mashhadi2021federated} presented an FL based beam selection scheme that utilizes on-vehicle Light Detection and Ranging (LIDAR) data. It was shown that the proposed scheme outperforms the previous state-of-the-art approaches in terms of classification accuracy and subsequently reduces floating point operations by a factor of 100. Moreover,
Qi \etal\cite{qi2020federated} used an asynchronous FL based transfer learning approach for pro-active handover in mmWave vehicular networks with varying mobility patterns. Their results show a reduction in uplink communication overhead and improvement in user's QoS. 
FedVCP is an FL based framework designed by Kong \etal\cite{kong2021fedvcp} for privacy-aware cooperative positioning in vehicular networks which was also demonstrated to outperform previous schemes. Furthermore, Saputra \etal\cite{saputra2019energy} implemented an FL based energy demand estimation method in a network of electric vehicles.
A selective FL aggregation scheme for vehicular networks was presented by Ye \etal\cite{ye2020federated} where only those local models are chosen for global aggregation which have high quality images and sufficient compute power.
\subsection{Network slicing}
Personalized network as a service has been considered as one of the main motivations for development of 5G networks. In heterogeneous networks, the aim of radio access network (RAN) slicing is to deliver tailored personalized services for mobile users with varying QoS needs. As a result, the most important aspect of RAN slicing is determining how to efficiently distribute network resources while still satisfying user QoS requirements. In the existing literature, there have been some works on application specific network slicing using deep learning \cite{du2018deep, 9348019, thantharate2019deepslice}. In application oriented slicing, the incoming traffic is analyzed at the packet level and each packet is classified into labels based upon the application using a neural network that was trained to learn the non-linear mapping between input and output data. Here input data consists of packet properties and sender device information, whereas output data depends on the application type. Based upon the classified packets, the network resources are distributed so as to satisfy the respective applications. Next generation wireless networks tend to have extremely dynamic usage patterns which are very tough to capture by supervised learning based resource allocation models. Deep reinforcement learning (DRL) has been known for its capability to make sequential decisions under dynamic constraints and thus has been considered in many works on network slicing \cite{li2018deep, qi2019deep, koo2019deep, shome2021deep}. Most of the above discussed methods require private data and computation intensive models which can be of serious concern when deployed in the real-world. As per 3rd Generation Partnership Project (3GPP) guidelines, network slices are required to be isolated from each other, which is not possible with the above discussed centralized learning schemes. These issues have been addressed in literature by designing FL based network slicing strategies. For instance, Brik \etal\cite{brik2020predicting} used FL to train a model that predicts key performance indicators (KPIs) of all the isolated running network slices in a distributed manner ensuring privacy among the slices. Their model was able to achieve equivalent performance in terms of mean squared error metric while consuming significantly lesser amount of network overhead. 

\subsection{Computation offloading at the Edge}
Offloading and caching computations at the wireless edge is a potential approach for reducing network traffic and backhaul burden while performing big data deliveries across the network. It has been shown that the same popular content is requested from the content delivery network multiple times by different users, and these duplicated requests result in unnecessary network latency. As discussed earlier, the Quality of Experience (QoE) in future wireless networks will be a complex metric jointly quantified by latency, bandwidth utilization, throughput and other KPIs. Edge computation offloading can provide a significant boost to the overall QoE by optimizing all the KPIs due to much lower network load.
A few works considered machine learning based content caching schemes using K-nearest neighbours\cite{8170936}, kernel ridge regression\cite{8711328}, bayesian learning\cite{8510864}, Collaborative filtering\cite{8951133}, but these traditional machine learning approaches fail to capture the real-time dynamics of content popularity at different locations and preferences among different users.
However, in a study\cite{mohammed2021performance}, it was shown that a  neural network outperforms traditional machine learning approaches for edge caching whenever the number of input features is high and communication range and file sizes are large. This shows that deep learning is a key enabler for wireless content caching and thus there has been a lot of work in literature on using deep learning for edge caching which proved to have better performance than conventional baselines\cite{8624176, 8576500, rahman2020deep, bhandari2021deep}. Reinforcement learning has also been used widely for edge content offloading due to its real-time decision making capabilities and directly being able to take actions to control the network\cite{8629363} and it has shown good results in performing personalized content caching based on user's preferences\cite{8737456}. For real-world deployment, the above mentioned methods may not be feasible due to the requirement for user's personal private data in traditional centralized learning architectures. Thus, intelligent wireless content caching models implemented with FL seems to be the best possible approach due to the privacy-preserved nature of the FL. 

To this end, Chilukuri \etal presented FedCache\cite{chilukuri2020achieving}, an FL based edge caching methodology that dynamically allocates the net available cache resource per edge node while ensuring a high cache hit ratio. Their system model assumes each edge node is able to divide its available cache space among all the different classes of data flowing in the network. They consider the real-time state of each class of data as a tuple with average inflow and outflow rates. During the initial phase, they collect data at the edge nodes with input features as state per data class and cache hit ratio as the output which needs to be predicted in real-time. Then, a synchronous FL phase starts at the node and the local models learn to predict the cache hit ratio based upon allocated cache space and network state. These local model's parameters are transmitted to the central PS where the global model learns from these aggregated parameters. The trained global model makes decision at the edge node of how much cache space to allocate for optimal wireless caching at that moment of time. These two phases keep on repeating at specific time intervals leading to a lifelong learning strategy which is able to adapt to the changing network dynamics. Experimental results show that FedCache has a better performance than chosen conventional methods such as Proportional split cache and unified cache. Furthermore, Shahidinejad \etal\cite{shahidinejad2021context} proposed an FL based context-aware approach for intelligent content caching at the wireless edge in a multi-user mobile edge computing (MEC) system. Their framework consists of monitor, analyze, plan and execute (MAPE) phases. In the monitoring phase, all the available network, resources, media and sensor information at the edge are collected and termed as context information. During the planning phase, DRL agents are trained to learn the content offloading policy from experience replay memory. As the DRL agents need a lot of data and compute to train and they require private data, it can lead to network overload as well as privacy concerns if trained on mobile devices. To tackle these challenges, the authors used FL based distributed training strategy for the DRL agents which led to a significant amount of energy savings than centralized approaches. 

\subsection{Internet of Things}
The broad use of smart technologies, cloud computing, and the  IoT has been accelerated by recent improvements in wireless communications\cite{jiang2020federated}. IoT is an emerging paradigm with the potential to realize the goal of smart cities\cite{khan20206g}. The massive scale of data generated by billions of IoT devices globally would pave a path for deploying intelligence at the edge. The major challenge is that current generation centralized learning schemes do not guarantee user privacy, and also are not able to handle such massive data in real-time. To cope with these limitations, FL would be a promising solution due to its inherent abilities to preserve privacy and to reduce network overload. 

Smart digital healthcare services are one of the most key applications in modern smart cities. FL has the potential to enable precision medicine on a broad scale, leading to models that provide impartial choices, optimally represent an individual's physiology, and are sensitive to uncommon illnesses while respecting governance and privacy issues\cite{rieke2020future}. FedHealth, the first federated transfer learning framework for wearable healthcare was developed by Chen \etal\cite{chen2020fedhealth} in an attempt to address these problems. FedHealth aggregates data via FL and then uses transfer learning to construct reasonably personalized models. In the real-world, labeled data necessary for supervised learning approaches is scarce, whereas unlabeled data is available at a massive scale. Keeping this fact in mind, Albaseer \etal \cite{albaseer2020exploiting} proposed FedSem, an FL framework which implements a semi-supervised learning approach in order to make use of the large amounts of unlabeled data available in smart cities and train optimal models with minimum supervision. Acquiring enough data in the medical imaging area is a big difficulty as labeling medical imaging data necessitates expertise of a medical specialist. Collaboration across institutions might address this difficulty, however transferring medical data to a centralized server raises a number of legal, privacy, technological, and data-ownership issues, particularly among multinational organisations. To this end, Sheller \etal \cite{sheller2018multi} provided the first use of FL for multi-institutional collaboration, enabling deep learning modelling without exchanging patient data to train efficient brain tumor segmentation models. Similarly, Liu \etal \cite{liu2021federated} presented an FL framework for collaboratively learning power consumption patterns in distributed smart grids in a privacy-preserving manner. Moreover,  Ta{\"\i}k \etal \cite{taik2020electrical} used Edge FL for household electrical load forecasting.

\section{Open problems and Future directions}\label{sec6}
\subsection{Hierarchical FL}
Hierarchical FL\cite{abad2020hierarchical} (HFL) is a framework that implements Edge FL across heterogeneous wireless networks with a novel three-level hierarchy which constitutes client devices, small base stations (SBS) and macro-cell base stations (MBS). For each SBS, a set of client devices are selected and the models are trained on each of these clients using FedSGD algorithm and then the parameter updates are sent to the respective SBS. The MBS and SBS communicate at periodic intervals so as to maintain a central model. Experimental results verified that HFL achieves better performance in terms of latency and accuracy than conventional FL. This concept can be generalized into multi-level hierarchy based FL and the architecture of the system can be designed based on particular use-case. The major benefit from HFL is due to much lesser computational overload on each of the distributed SBSs compared to only one MBS present in traditional FL. Most privacy enhancing methodologies have a high time complexity as well as computational/network overload. This reduced overload due to HFL opens room for improving privacy by utilizing these methodologies\cite{wainakh2020enhancing}. Moreover, traditional FL relies on a single PS which may lead to an entire FL system crash under emergency situations/attacks. HFL helps to overcome the limitation due to the presence of multiple aggregation servers instead of one.
\subsection{Decentralized FL}
Traditional FL schemes involve a central PS where the aggregation of local model weights takes place. Unfortunately, the PS may be the victim of malicious attacks or system failures which may lead to a major downfall in FL performance and in the worst case, the entire distributed training can come to a halt. Also, the client devices and the centralized server need to transmit model updates between each other through wireless links multiple times, which leads to a large network overhead when a huge set of client devices participate in FL training. Taking these facts into account, Khan \etal \cite{khan2020dispersed} presented a framework termed 'Dispersed FL' which trains a global FL model in a fully decentralized manner and compared it with hierarchical FL. The authors mentioned that this fully decentralized approach has some serious issues such as increase in latency and extra bandwidth requirements for transmitting local and global models during the training phase among the distributed nodes. Chen \etal\cite{chen2020wireless} proposed Collaborative FL (CFL) framework which is a hybrid approach where a set of client devices may be connected to the central PS whereas other client devices connect with nearby devices to transmit parameters after each local training step. Considering the fact that mobility of distributed devices / nodes has not been taken into account in any of the existing works, as a future work, researchers can investigate the impact of mobility on the performance of FL. 
\subsection{Federated personalization}
Local datasets on client devices have different varying features correlated with the personal preferences and characteristics of the user. For example, in the case of computational offloading, different users would have different usage patterns and it would be very beneficial to use those patterns such as app-wise screen time, choices of photos and search queries etc., as input features to the local models as these can make the models much more efficient in understanding what the user actually wants. When those local gradients would be aggregated into the global model, the global model would learn what all the participating users personally prefer. In Hierarchical FL, this kind of personalization can be more useful in cases where geographically nearer users would have certain features which are local to them and do not benefit users from other areas. For instance. in case of offloading video content, preferences are correlated with locations.
Thus, as a future research direction, we recommend searching for local data which can be used as input for capturing personal features for FL.

\subsection{Security}
In spite of the fact that, a privacy-preserving feature is immanent in FL process, it can still be vulnerable to numerous security threats and privacy issues mainly due to involvement of many clients. There are relatively few articles in the literature that identify dangers of the vulnerabilities and propose methodologies for mitigating those dangers \cite{mothukuri2021survey}. For the iterative learning process of FL to achieve higher accuracy, the PS has to communicate extensively with a large number of clients. However, the probability of having insecure communication links also increases with increasing the number of clients, thus an accuracy-privacy tradeoff may exist, and we suggest investigating this important tradeoff in future works.
Poisoning is a major category of attacks that use malicious data in the training phase, or adversely update the model parameters in order to significantly degrade performance of the FL learning model. Additionally, Generative Adversarial Networks (GANs) have the power to launch attacks against any machine learning model, including the FL model. It is shown in \cite{hitaj2017deep} that these GAN based attacks have high severity because such attacks have an ability to steal information of any client by generating prototypical samples of the targeted training dataset.
There have been some efforts in literature to tackle such security challenges. Differential Privacy based perturbation approaches \cite{dwork2006calibrating} conceal certain sensitive attributes, thereby making the data impossible to restore, consequently safeguarding user privacy. Secure multi-party computation (SMC) is another popular technique that employs a four-round interactive protocol that can be enabled during the reporting phase of a given FL communication round \cite{zhu2020relationship}. After receiving messages from each device in a protocol round, the server returns an individual answer to each device. The third cycle is referred to as the "commit phase" in which devices send encrypted model updates to the server. There is a final stage where devices give enough cryptographic secrets so that the server may unmask the aggregated model update. Another popular technique is Homomorphic encryption (HE) \cite{aono2017privacy} which is a key-based security method. HE enables computations on encrypted data. By leveraging HE in the context of FL, first the participating clients create public and private keys, which are used to encrypt locally learned models. This is followed by a secure server-side aggregate of all model updates received from clients. The global model changes are decrypted by the clients using the private keys. It is quite evident that the increased security is provided by using HE, however, the computational complexity of the cryptographic operations adds additional overhead on the resource-constrained clients, in terms of cost, time and power consumption. Moreover, these encrypted models are significantly larger, thus also scaling up the overall communication overhead. Hence, one of the important future research directions is to develop robust security enhanced frameworks for FL with a focus on reduction in computational and communication overheads. In traditional communications, cryptographic operations are completely avoided by invoking approaches that are based on the principles of the physical layer security (PLS) \cite{Waqar_TVT}. As PLS may serve an alternative to cryptographic security techniques even for the FL, PLS based approaches should be worthy enough to be considered in future for the FL.

\section{Conclusion}
An ever-growing demand for data hungry applications on a wireless network along with the phenomenal success of Deep Learning in various fields has resulted in a natural integration of the two technologies, namely, deep learning and wireless communications. However, the rising concern of private data leakage in conjunction with the high latency associated with uploading the entire raw data to a central cloud, makes the centralized deep learning techniques unattractive. To this end, recently, FL has been proposed and owing to its privacy-preserving and communication efficient features, it has become a natural choice for the next generation communication systems. However, there exists an interdependency between the two technologies as performance of the FL strongly depends on conditions of the wireless channels. Therefore, the aim of this survey is to excavate this bidirectional relationship between FL and wireless communications in a holistic manner. As a result, we provide a comprehensive survey of the techniques that have been proposed in state-of-the-art literature for the communication efficient FL, wireless communications for FL and FL for wireless communications. Moreover, this is the first survey article that discusses the role of the smart radio environments for the performance improvement of the FL. Lastly, we also discuss the open problems for the FL and provide future directions which will be helpful for those researchers working in the intersection of the two emerging paradigms i.e., FL and next generation wireless communications.

\bibliographystyle{IEEEtran}
\bibliography{wileyNJD-AMA}

%




\end{document}